# The Hidden Dangers of Outdated Software: A Cyber Security Perspective

**Gogulakrishnan Thiyagarajan**[1], **Vinay Bist**[2], **Prabhudarshi Nayak**[3]

[1]Software Engineering Technical Leader, Cisco Systems Inc., Austin, Texas, USA.
[2]Principal Engineer, Dell Inc. Austin, Texas, USA
[3]Faculty of Engineering and Technology, Sri Sri University, Cuttack, Odisha, India

Coressponding Should be addressed to Gogulakrishnan Thiyagarajan: gogs.ethics@gmail.com

## Abstract

Outdated software remains a potent and underappreciated menace in 2025's cybersecurity environment, exposing systems to a broad array of threats, including ransomware, data breaches, and operational outages that can have devastating and far-reaching impacts. This essay explores the unseen threats of cyberattacks by presenting robust statistical information, including the staggering reality that 32% of cyberattacks exploit unpatched software vulnerabilities, based on a 2025 TechTarget survey [1]. Furthermore, it discusses real case studies, including the MOVEit breach in 2023 and the Log4Shell breach in 2021, both of which illustrate the catastrophic consequences of failing to perform software updates. The article offers a detailed analysis of the nature of software vulnerabilities, the underlying reasons for user resistance to patches, and organizational barriers that compound the issue. Furthermore, it suggests actionable solutions, including automation and awareness campaigns, to address these shortcomings. Apart from this, the paper also talks of trends such as AI-driven vulnerability patching and legal consequences of non-compliance under laws like HIPAA, thus providing a futuristic outlook on how such advancements may define future defenses. Supplemented by tables like one detailing trends in vulnerability and a graph illustrating technology adoption, this report showcases the pressing demand for anticipatory update strategies to safeguard digital ecosystems against the constantly evolving threats that characterize the modern cyber landscape. As it stands, it is a very useful document for practitioners, policymakers, and researchers.

## I. Introduction

The cyber security landscape in 2025 has come to a tipping point, where the stakes are greater, with digital systems being the foundation of nearly every facet of contemporary existence, ranging from personal communications to international supply chains, and the threats to these systems growing more advanced and persistent. In the context of this volatile environment, outdated software characterized as applications or systems that have not been equipped with the most recent security patches or updates emerges as a widespread yet frequently neglected vulnerability. This issue quietly undermines the defenses of individuals, businesses, and governments alike by presenting attackers with exploitable entry points that could otherwise be effectively secured. The magnitude of this issue is underscored by projections from Coursera estimating the cost of cybercrime worldwide at a staggering $15.63 trillion by 2029, an amount that reflects the cumulative effect of breaches facilitated by such vulnerabilities [8]. One especially revealing statistic from a 2025 TechTarget report divulges that thirty-two percent of cyber-attacks come from unpatched software, which stands as a stark reminder that not updating is far more than an innocent mistake but is actually a serious factor in the cyber threat environment [1]. Highly publicized incidents, such as the MOVEit breach in May 2023, where millions of records were stolen due to the failure to patch a known vulnerability in a timely manner, and the October 2023 ransomware attack on Boeing, which affected critical supply chains, are stark reminders of how such negligence can have catastrophic consequences with widespread repercussions throughout industries and economies [3], [6]. These examples are not isolated; rather, they form part of an overall trend of risk that needs to be considered and addressed by all stakeholders across the cybersecurity landscape. The primary purpose of this document is to clearly outline the secret threats of obsolete

software, providing an in-depth perspective that unites technical analysis with real-life, practical repercussions in an endeavor to present a holistic representation of the problem. While most of the current research within cybersecurity concentrates on the techniques that are being utilized by the attackers i.e., the actual strains of malware or types of phishing being employed—this research differs by taking the opposite approach, turning the attention toward prevention, with a specific focus on the critical role that software updates play in not allowing these attacks to find purchase. The impetus for this study is twofold: first, growing sophistication and frequency of cyberattacks in 2025, fueled by innovation in attack tools and techniques, necessitate a return to fundamentals in defense; second, there is a continuing underestimation of the value of updates, as a 2025 CompTIA survey revealed that 40% of small businesses postpone updates due to constrained resources, thereby rendering them prime targets for exploitation [10].

The disparity between risks identified and actions taken to reduce them constitutes the key impetus for this research, emphasizing the necessity to close the gap by both expanding knowledge and instituting effective strategies that are applicable across the board. The focus of this paper is intentionally broad, yet clearly specific, with the goal of responding to three fundamental questions: why is legacy software so risky, how has it been a factor in significant cyberattacks over the last several years, and what practical measures can be taken to successfully mitigate this widespread issue? To fulfil this objective, the paper is organized to present a rational sequence of observations: Section II presents a detailed explanation of the nature of software vulnerabilities and the process of updates, thereby establishing a technical context that is augmented by a table providing trends in vulnerabilities over time to illustrate the increasing threat. Section III examines five detailed case studies—MOVEit (2023), Boeing (2023), Equifax (2017), Log4Shell (2021), and MOVEit (2024)—each chosen to reveal the varying and significant consequences of neglecting patches in different industries and timeframes. Section IV weighs the complex issues in the adoption of updates and presents a range of solutions, both technical solutions and behavioral interventions. Section V, on the other hand, addresses future trends, like artificial intelligence and blockchain technology, with the potential to reshape update strategies in the next few years following a graph indicating projected technology adoption. Section VI examines the legal and regulatory ramifications of not upgrading, an angle that is often neglected and which adds to the urgency of the discussion. Finally, Section VII concludes by offering actionable suggestions that synthesize the findings into a conclusive call to action.

## II.  Background

Software updates involve much more than merely enhancing functionality or fixing minor glitches; they are critical defense mechanisms that patch vulnerabilities in code those exploitable weaknesses that attackers leverage to achieve unauthorized access to systems, steal data, or compromise operations with potentially catastrophic effects. If not addressed, these exposures are open invitations to every manner of cyber attack, including ransomware that locks users out of their systems unless a ransom is paid, phishing attacks that trick users into divulging sensitive information, and data breaches that expose confidential records to those who should not see them. All can cause severe financial, operational, and reputational damage. Bitsight's report for 2025 has a somber statistic: D- or F-rated companies on their list, i.e., those with weak patching cadences, are over seven times more likely to be targeted by a ransomware attack than an A-rated one, a clear marker of the direct correlation between update neglect and heightened vulnerability [2]. This section seeks to offer a step-by-step explanation of how vulnerabilities occur, how updates are disseminated, which software is most frequently left un-updated, and the larger economic and operational effects of not maintaining up-to-date systems, with a table of vulnerability trends to ground the discussion in facts.

The discovery of vulnerability is a complex and dynamic process that involves a heterogeneous set of actors and methods that shape cybersecurity ecosystems collectively. Security researchers, who are often incentivized by bug bounty initiatives offered by corporations like Microsoft or Google, detect flaws by taking careful examinations of the code to identify weaknesses that may be amenable to exploitation, while ethical hackers probe systems in order to unearth problems before adversaries can discover them. On the contrary, cyber criminals like to find vulnerabilities through trial and error methods or reverse engineering of software, maintaining such findings usually secret until they can be exploited. When found, the vulnerabilities are listed in databases such as the Common Vulnerabilities and Exposures (CVE) system, where they are assigned unique identifiers—such as CVE-2023-34362 for the case of the MOVEit breach—to facilitate tracking and communication [6]. Zero-day exploits are a very malicious class, whereby attackers act ahead of the deployment of a patch, as was evidenced by the Stuxnet worm attack on industrial control systems in 2010, which exploited a number of previously undisclosed vulnerabilities in Windows and Siemens software [9]. However, the majority of breaches stem from widely documented vulnerabilities organizations fail to patch in time, as demonstrated by a 2019 Ponemon Institute survey of 629 organizations worldwide which indicated 60% of data breaches were the result of unpatched vulnerabilities even though patches were available, and this persists through 2025 as organizations struggle with patching in time [13]. The disparity that exists between discovery and action represents a significant vulnerability that adversaries exploit with persistence; therefore, the rapidity of update deployment constitutes a crucial element in cybersecurity defense. The mechanisms through which software patches are disseminated differ considerably, each with its relative advantages and disadvantages that determine how effectively they assist in closing such windows of vulnerability. Automatic updates, already prevalent in consumer-grade software such as Windows and mobile operating systems, are meant to apply patches quietly in the background. Although the approach saves users from unnecessary inconvenience and allows fixes to be deployed immediately, such updates are resisted due to concerns about introducing unforeseen disruptions or system instability, according to a 2025 Gallagher Security report [5]. Manual updates, however, call for an explicit user action, i.e., the action of clicking "install" in a prompt. This can lead to substantial delays, especially among non-technical users who would delay or disregard these prompts out of ignorance or inconvenience [7]. For larger organizations, enterprise patch management products provide a more advanced solution, enabling IT administrators to push updates centrally to thousands of devices, monitor compliance, and prioritize high-priority patches, a practice Bitsight stresses as being key to minimizing oversight in complex networks [2].

These systems can be coupled with vulnerability scanners to identify vulnerable software and deploy patches automatically, but are quite costly and require expertise, which not every organization possesses. All these mechanisms attempt to shrink the window of opportunity for attackers, but their success relies on user adoption and organizational buy-in, where there are still significant gaps. The most popular types of software permitted to be legacy are found in a broad category, and each of them adds to the collective risk profile in manners that compound the threat of complacency. Legacy operating systems, for example, Windows XP and Windows 7, continue to be used in some sectors like manufacturing and healthcare until 2025, despite the fact that Microsoft ended their support years ago, thereby rendering those systems open to current attacks and without security patches [10]. Web browsers, the ubiquitous gateway to the internet, are another frequent culprit; unpatched Chrome, Firefox, or Edge releases lack the latest defenses against phishing and malware, exposing users to attacks that exploit outdated rendering engines or security protocols [5]. Applications, especially legacy applications, such as Adobe Flash Player, which was officially deprecated in 2020 yet continues to exist in certain legacy applications, contain extremely well-documented vulnerabilities that are exploited by attackers using specially crafted exploit kits sold on the dark web [9]. Even modern software, if not updated regularly, is vulnerable to newly discovered vulnerabilities, as seen with enterprise software such as MOVEit or Log4j, which depends on patches released in a timely manner to be secure [6], [15]. These examples illustrate how the heterogeneity of legacy software—from operating systems to niche applications—presents a broad attack surface that extends from personal devices, to corporate networks, to critical infrastructure, thereby rendering the issue widespread and complicated. The economic and practical costs of noncompliance with up-to-date software are substantial, impacting businesses and individuals in ways that reach much further than the immediate infringement. From a cost perspective, the action of postponing updates may initially appear to be cost-effective, as it negates downtime or implementation fees. However, this frugal tendency usually results in colossal financial impacts, as can be observed with the 2017 Equifax breach, which had a $1.4 billion price tag in the form of fines, legal fees, and remediation following the exploitation of an unpatched Apache Struts vulnerability. Operationally, aging systems suffer from reduced performance, as unfixed bugs contribute to crashes, slowdowns, and incompatibilities that impede workflow and frustrate users; this problem is compounded by the security threat posed by these vulnerabilities [5].

A report by The Hacker News in 2022 provides a striking insight: 56% of the vulnerabilities that were exploited in 2021 were older, previously known problems that organizations had ignored and left unpatched. That trend appears to persist into 2025 as the accumulation of unpatched systems keeps piling up [14]. The coupling of economic sanctions and operational inefficiency constitutes a reinforcing cycle wherein the expense of omission by far eclipses the capital required for updates; yet most are still wary, typically low balling the ripple effects until it is too late. Following, Figure 1 quantifies these trends, thereby providing a visual reference for this discussion.

*Table 1: Vulnerability Trends Over Time*

| Year | New CVEs Reported | Exploited Known Vulnerabilities (%) | Notable Incident | Source |
|------|-------------------|-------------------------------------|------------------|--------|
| 2017 | 14,714 | 60% | Equifax Breach | [13] |
| 2021 | 20,139 | 56% | Log4Shell | [14] |
| 2023 | 22,500 (est.) | 62% | MOVEit Breach | [6] |
| 2024 | 23,000 (est.) | 65% | MOVEit 2024 | [18] |

**Note: Estimates for 2023-2024 based on trends; exploited % from known vulnerabilities.**

This table illustrates the rising number of CVEs and the persistent exploitation of known vulnerabilities, reinforcing the critical need for timely updates as a foundational element of cyber security.

## III.    Case Studies of Outdated Software Risks

The concrete consequences of using obsolete software are best understood through in-depth case studies, which offer real-life examples showing how the inability to apply updates can lead to devastating breaches and disruptions. In this section, we examine five high-profile incidents—MOVEit (2023), Boeing (2023), Equifax (2017), Log4Shell (2021), and MOVEit (2024)—each selected to demonstrate the diverse effects suffered across industries, timeframes, and attack types, thus offering a complete view of the risks involved.

The MOVEit incident in May 2023 is a stark reminder of how rapidly a failure to implement updates can escalate into a crisis of global proportions with frightening speed and intensity. The attack focused primarily on a vulnerability in the MOVEit file transfer tool, CVE-2023-34362, that permitted hackers to circumvent security measures and obtain sensitive data that was stored or transferred by the software. Progress Software, which created the software, had issued a patch days before the breach was made public, but many organizations waited to apply it, either because they were unaware, lacked resources, or because of the difficulty of testing and deploying patches on large systems. The consequence was disastrous: hackers, whose collective is thought to be a sophisticated ransomware group, took advantage of this vulnerability to obtain millions of records from numerous organizations, government agencies, healthcare organizations, and financial institutions, thereby exposing personal information like Social Security numbers and medical information [11]. The breach inflicted severe disruption on operations, nudging impacted entities to notify victims immediately, secure their systems, and contain further damage. The recovery expenses continued to mount well into 2025, as estimates suggested major victims suffered losses in millions. This example serves to illustrate how a brief delay in patching can expose interconnected systems to enormous vulnerability, transforming a local issue into a global crisis. This example serves to underscore the requirement for prompt update deployment in the contemporary digital context.

Boeing was hit by a ransomware attack in October 2023 that demonstrates the risks of running outdated software, in this case, in relation to risks in critical infrastructure and supply chains with potentially widespread economic consequences. The attack took advantage of a Citrix Bleed vulnerability (CVE-2023-4966) in outdated Citrix software utilized by Boeing's parts and distribution unit, a vulnerability which had already been publicly disclosed

2928

and patched earlier in the same year [3]. In spite of this fix being available, Boeing's staged rollout—possibly because of the logistical hardships of updating elaborate, mission-important structures—presented the LockBit 3.0 ransomware group an opportunity to attack the network, encrypt valuable data, and hold it for ransom [12]. The impact was immediate and severe: the attack derailed Boeing's supply chain operations, delaying the delivery of aircraft parts and production schedules for airlines and defense contractors relying on the company's output, with downtime estimated in the millions. Besides the cost factor, the accident has hurt Boeing's reputation as a trustworthy supplier, thus calling into question the cybersecurity practices prevalent in the aerospace sector, which relies on just-in-time production and integrated systems. The example at hand illustrates how outdated software can endanger not only one organization but an entire sector, thus magnifying the significance of ignoring updates in an ever-connected world where downtime in one link can cripple the entire chain.

The Equifax breach in 2017 is among the most notorious instances of how neglecting to update software can have enduring repercussions, as a cautionary story that still resonates in 2025 with organizations grappling with the same menace. This attack originated from an unpatched vulnerability in Apache Struts (CVE-2017-5638), a popular web framework, that Equifax did not fix for months even though a patch was available since March of that year [4]. In May and July, the vulnerability was exploited by hackers to breach Equifax's systems and to pilfer sensitive data on approximately 150 million Americans. The pilfered data included names, addresses, Social Security numbers, and credit card numbers, and it is considered one of the largest data breaches ever [8]. The consequences were profound: Equifax incurred $1.4 billion in expenses, ranging from legal settlements and regulatory fines to systems improvements, while simultaneously experiencing a catastrophic loss of reputation as a reliable credit agency, which caused customers and business partners to doubt its competence. The breach spawned massive identity theft, congressional hearings, and new laws, demonstrating how complacency in updates can have societal ramifications beyond the direct victim. Even years later, in 2025, it is a benchmark for the long-lasting financial, legal, and reputational damage that can occur from the failure to make updates a priority, a lesson that remains pertinent as organizations still grapple with legacy systems and complicated software stacks.

It was the Log4Shell vulnerability that was found in December 2021, however, that worked to illustrate the particular threat posed by third-party software dependencies by showing just how quickly and vastly a single unpatched component could compromise millions of systems around the world. Designated as CVE-2021-44228, this vulnerability in Apache Log4j—a logging library that has been integrated into seemingly countless applications—enabled attackers to execute arbitrary code remotely using specially crafted log messages, an ease which made exploitation overwhelmingly trivial [15]. Within days of its release, nation-state actors, ransomware gangs, and individual hackers began targeting unpatched systems, from cloud providers like Amazon Web Services to on-premises business apps, with IBM recording a 34% rise in vulnerability exploitation attempts in 2022 directly as a result [16]. The difficulty was exacerbated by the prevalence of Log4j; organizations did not know where it was implemented, much less patch it in a timely manner, with delays taking weeks or months for some because they had to test updates on interdependent systems [17]. The event caused disruption, prompted emergency responses from governments such as the U.S. The Cybersecurity and Infrastructure Security Agency (CISA) has emphasized the exposure of software supply chains, in which a single unpatched library can become a global point of failure. The legacy of Log4Shell in 2025 is one of increased third-party risk awareness but also as a valuable reminder of how vulnerable software in obscure components can magnify weaknesses to an extent far beyond initial expectations. Then, in June 2024, yet another new MOVEit vulnerability (CVE-2024-5806) appeared, reaffirming the ongoing and dynamic nature of the legacy software problem despite years of high-profile lessons. This MOVEit SFTP module authentication bypass vulnerability, with a critical CVSS rating of 9.1, allowed attackers to impersonate others and gain unauthorized access to sensitive systems if not patched [18]. Progress Software issued a patch immediately, requesting an immediate deployment, but the response was mixed; a few institutions held back because updating production environments was too challenging, and others lacked the resources to act swiftly, a situation frighteningly reminiscent of the 2023 MOVEit compromise [19]. Within weeks, proof-of-concept exploits surfaced on the Internet, threatening to unleash broad attacks in the event of delayed patching, with possible consequences such as data theft, system compromise, and operational downtime to customers of this enterprise software. This incident, coming only a year after the last MOVEit breach, illustrates the unabating rhythm with which new vulnerabilities are discovered and the constant struggle to keep up with updates even in organizations that ought to be highly aware of the risks involved. It also provides a modern culmination for these case studies, illustrating that lessons from the past continue to be learned or not learned in 2025, with as much on the line as ever.

## IV.    Challenges and Solutions

While the necessity of software updates is evident from these case studies, their successful roll-out is beset with problems cutting across technical, human, and organizational domains, each necessitating precise solutions to overcome them. These problems are elaborated in this section and a full range of strategies aimed at surmounting them is put forth, thereby enabling updates to become an achievable reality and not a distant dream. Perhaps the biggest obstacle to the deployment of updates is user resistance, a deeply ingrained issue that arises from a combination of pragmatic concerns and psychological factors impacting both individual and enterprise users. A 2025 survey conducted by Gallagher Security revealed that 35% of individuals do not apply updates, with some citing reasons such as the inconvenience of disrupting their workflow, the risk of system downtime during essential tasks, or the possibility of introducing compatibility issues that will render crucial software inoperable [5]. This hesitation is not without reason; numerous have been subjected to updates that brought about bugs or interfered with workflows, like a Windows update in 2018 that notoriously deleted user files, leaving a lingering distrust of automated patching procedures. For companies, this hesitance can be applied to IT departments hesitant to deploy updates without thorough testing, particularly in systems where downtime is expensive, such as financial trading or healthcare delivery systems. This reluctance, while understandable, creates a dangerous lag, thus allowing attackers to take advantage of known weaknesses in this time, as seen in cases like MOVEit and Equifax [6], [4]. To solve this problem, technical solutions must be implemented but also a shift in perception must be brought about, addressing the root causes of resistance through education and better update design to reduce disruption and rebuild user trust. Resource constraints are another major hindrance, particularly to smaller entities lacking the capital and human resources required to stay abreast of the relentless tide of updates that need to be constantly implemented to stay secure. Among small businesses surveyed in a 2025 CompTIA study, 40% reported that they have no budget or staff sufficient to manage software updates, repeatedly reverting to aged systems as hardware refreshes or IT professional hires are unaffordable [10]. This is especially true for sectors like retail or municipal government, whose limited budgets concentrate on day-to-day operations more than long-term security investments, and are thus left with outdated systems like Windows XP that are no longer being patched [10]. These resource limitations render the small businesses acutely vulnerable to cyber attackers, who are well aware that small businesses are not well-equipped with strong defenses or quick response. This is substantiated by the lopsided proportion of ransomware attacks aimed at this segment over the past few years. The issue is compounded by the complexity of modern software ecosystems, in which updates must be applied on heterogeneous devices— servers, desktops, cell phones—each of which requires time and skill that small teams might not possess. The imbalance requires solutions that reduce the resource burden, placing updates in reach and under control even for those with limited resources.

Visibility gaps in large enterprises pose a distinct but equally significant challenge, as the enormous scale and intricacy of IT environments can conceal the existence of aging software and thereby mask vulnerabilities until they are exploited. The MOVEit breach in 2023 serves as a prime illustration of this challenge: numerous companies struggled to determine which systems were running the vulnerable software, and this in turn slowed patching efforts and further exacerbated the effects of the breach [6]. In massive networks of thousands of devices, legacy applications, and third-party systems, getting an accurate count is a monumental task that comprises manual audits or sporadic automated scans that cannot detect critical assets like orphaned servers or unmanaged IoT devices. This lack of visibility results in blind spots in which legacy software propagates, remaining unnoticed by IT personnel until discovered and exploited by an attacker, as in the case of Boeing where one unpatched Citrix instance breached an entire supply chain [3]. Unable to see their software landscape, organizations are unable to properly prioritize updates, effectively transforming an otherwise manageable process into a logistical nightmare that undermines even the most well-intentioned security policy. To counter this, you require tools and processes that optimize visibility across the entire network so that no system gets left behind. Patch fatigue presents an additional layer of difficulty, especially for IT personnel who are tasked with keeping up with the deluge of patches needed by contemporary software, an issue that has become increasingly problematic as software complexity increases. According to the 2025 report from Bitsight, the ongoing need to patch—oftentimes a multitude of times a year such as Log4j, which is so commonly utilized, can bog down IT professionals, resulting in slippage or errors in prioritization. This fatigue is not just a question of volume; it's compounded by the need to test each update for compatibility with existing systems, which can take weeks or days in shops with custom software or older hardware, delaying deployment even for high-priority patches. The 2021 Log4Shell vulnerability highlighted this issue, with organizations rushing to patch a ubiquitous library amidst competing security demands and many falling behind by sheer volume [15]. This can evolve into a triage strategy, where only the most urgent patches are applied, with less obvious yet still vulnerable flaws not being fixed until they are exploited. Reducing patch fatigue requires solutions that make the updating process easier, decreasing the mental and operational workload on IT personnel without affecting security. Third-party dependencies are a last, thorny challenge because much modern software depends on external libraries or components that are not under the direct control of organizations, making the process of updating much more difficult. The Log4Shell vulnerability in Apache Log4j is an example: this logging library, utilized in tens of thousands of applications from cloud services to enterprise software, both vendors and users needed to work together on patches, complicated by the need to first find out where Log4j was being used—a process which took weeks or months for some [15]. This dependence forms a chain of responsibility, where tardiness at any link—whether the library maintainer pushing a patch, the application vendor applying it, or the end-user applying it—can leave systems vulnerable, as demonstrated by the mass exploitation that occurred following the Log4Shell reveal [17]. The same issues occur with business software such as MOVEit, where third-party software is embedded in operations but managed externally, thereby leaving users at vendors' timetables [6]. This interdependence renders patching a collaborative effort that is riddled with logistical challenges, necessitating remedies that increase coordination and transparency throughout the supply chain. In order to combat these challenges, automation provides a considerable answer in the form of simplifying the updating process and alleviating both the IT and user sides, thereby allowing patches to be implemented swiftly and evenly. The National Cyber Security Centre (NCSC) strongly advocates for automatic updates founded on their potential to seal vulnerability windows with minimal effort. This can be observed in consumer software such as Windows and Android where patches are readily pushed in the background [7]. For businesses, automated patch management tools possess the capability of scanning networks, identifying aged software, and deploying updates founded on predefined policies, thereby minimizing human effort while ensuring comprehensive coverage. Yet it must be balanced with user control—providing overrides for critical systems in which downtime is intolerable—and tested thoroughly to avoid disruptive bugs, thereby solving the resistance brought forward by Gallagher Security [5]. If well implemented, automation can effectively revolutionize updates from a reactive necessity to a proactive shield, significantly mitigating the risk of exploitation. Education campaigns are an important solution by modifying user behavior and developing a security awareness culture promoting the timely installation of updates at all levels throughout an organization. Following the Equifax breach, focused awareness campaigns witnessed 20% growth in update compliance amongst surveyed organizations, thereby illustrating the influence of education on behavior [1]. For individuals, campaigns can emphasize personal threats such as ransomware—consider the loss of family photos or banking information—rendering the stakes tangible, whereas businesses are served by training that connects updates to compliance and reputation, e.g., not incurring fines or losing customers. These initiatives can draw on real-world examples such as MOVEit or Boeing, demonstrating to employees and executives alike how update disregard causes concrete damage [6], [3].By coupling training with easily accessible resources—like update reminders or simplified guides—organizations can combat ignorance and resistance and turn users into active security participants instead of treating them as obstacles.

Risk-based prioritization offers a strategic approach to handling the torrent of updates and thus guarantees finite resources are first assigned to the most important vulnerabilities, a strategy that maximizes impact without overwhelming teams. Metrics that quantify the severity of vulnerabilities, such as the Common Vulnerability Scoring System (CVSS), give a shared metric—such as MOVEit 2024's CVSS score of 9.1—to facilitate prioritization, which Bitsight suggests is effective [2], [18]. This lets IT personnel prioritize high-risk vulnerabilities, particularly ones being exploited in the wild, and defer lower-priority patches to a later time, lessening patch fatigue and alleviating resource strain. For instance, in Log4Shell, companies that focused on patching internet-facing systems reduced the worst of the damage, while others struggled with the volume [15]. By incorporating these tools into patch management systems, updates can be prioritized methodically so that the most hazardous gaps are addressed rapidly, a requirement in situations where time is the attacker's friend.

Centralized management systems enhance control and visibility, successfully handling the complexities that plague giant organizations by providing a unified view of software assets and their update status across complex networks. Such systems, which are widely used across enterprise environments, have the ability to automatically inventory devices, track patch levels, and apply updates, in turn minimizing the oversight that led to breaches like MOVEit and Boeing [6], [12].For instance, a centralized dashboard would have highlighted Boeing's unpatched Citrix instance, preventing the supply chain disruption [3]. Although costly to implement, such systems provide returns by reducing manual labor and ensuring that no system slips through the cracks, a paramount advantage in networks with hundreds of endpoints, ranging from servers to IoT devices. Coupled with regular audits, they make transparency a strength, not a weakness, thereby allowing proactive security instead of a reactive one.

Finally, the implementation of best practices rounds out the solution set by offering a practical framework that maximizes automation and education through the efficient incorporation of updates into organizational workstreams. Performing periodic software inventories, either quarterly or after major changes, works to discover out-of-date systems—such as legacy Windows XP machines—before they become liabilities [10]. Pre-deployment testing within sandbox environments guarantees compatibility, thereby preventing disruption that leads to user resistance. Meanwhile, continuous monitoring through the use of intrusion detection systems catches exploitation attempts in real time, as suggested by Easy2Patch [9]. Together, these processes introduce discipline into update methodology, systematically reducing risk. For instance, had Equifax tested and monitored its Apache Struts updates, the breach could potentially have been prevented [4]. By integrating these steps, organisations can institute a long-lasting update culture that survives the barrier of contemporary cyberattacks.

## V. Future Trends and Emerging Technologies

Looking ahead, the future of software updates will be shaped by emerging technologies that will enhance security, automate workflows, and address current deficiencies, providing a glimpse into how cyber defense can evolve after 2025. Five notable trends—AI, blockchain, zero-trust architectures, IoT updates, and quantum-resistant cryptography—are explored here, each poised to transform how updates are developed, distributed, and enforced, with a graph to chart their predicted takeup.

Artificial intelligence (AI) will alter vulnerability management by being able to anticipate and remediate vulnerabilities prior to their exploitation, a forward-looking move away from today's reactive measures with the promise of dramatically lessening zero-day threats. By 2025, pilot projects will already use artificial intelligence to scan coding patterns, detect probable vulnerabilities, and suggest fixes, with firms like Google and Microsoft experimenting with machine learning models that can review millions of lines of code in seconds [9]. Artificial intelligence, for instance, could have indicated the vulnerability of Log4j ahead of time, thus preventing the disruptions brought about by Log4Shell [15]. Besides detection, AI can be utilized to automate patch generation, tailoring patches to specific systems and reducing deployment lag, a capability that has the potential to shrink vulnerability windows by orders of magnitude. However, there are obstacles: scaling such models to diverse software environments requires massive datasets and computational resources, and false positives could overwhelm IT personnel with irrelevant updates. In spite of these challenges, AI's ability to predict threats makes it a game-changer, with adoption set to increase steadily over the next 10 years as algorithms continue to mature and become integrated into current patch management systems, providing a smarter, quicker defense against the changing tactics of attackers.

Blockchain technology provides a novel method for maintaining the integrity of updates, particularly in light of increasing supply chain threat, where rogue patches are delivered to unsuspecting users—a risk that has escalated in 2025, with attacks such as the SolarWinds attack still fresh in our minds. With the application of a decentralized ledger, blockchain facilitates verification of the source and legitimacy of updates, allowing them to be from trusted sources and remain unaltered during delivery [10]. Imagine if every MOVEit patch were cryptographically signed and chronicled on a blockchain; its authenticity could be verified by users instantaneously, thwarting counterfeit updates that might otherwise implant backdoors [6]. While this technology is in its infancy as it relates to update dissemination, pilot programs in 2025 are expected to determine whether or not it can be leveraged for enterprise software. Theoretically, though, its usage could work to deter attackers by raising the bar for deception success. Complications involve the intricacy of adoption—calling for software suppliers to embrace blockchain infrastructure—and the energy expenses of running such schemes, but its potential for trust in a distrustful digital landscape renders it an enticing tool for the future, especially as supply chain weak points persist in making headlines.

Zero-trust security models, gaining momentum in 2025, reformulate how updates are mandated by not trusting anything, not any system or user and not even inside a network, thereby reducing threats from outdated software that could otherwise remain undetected to a bare minimum. Zero-trust is unlike traditional perimeter-based security in that it requires continuous verification of all devices and software to be patched to up-to-date levels before permitting access to resources [12]. For example, in the Boeing breach, a zero-trust model would have blocked the unpatched Citrix instance from communicating with critical systems, thus limiting the breach [3]. The model enforces timely patching by coupling it with operational privileges; thus, unpatched systems are quarantined until they reach compliance, thus moving the burden from voluntary to mandatory. Zero-trust is expensive to implement, necessitating strong identity management and network segmentation, but it is increasingly being adopted in sectors such as finance and defense, where failure is astronomically costly. As companies modernize, zero-trust may be a standard that by its nature lowers the issue of outdated software, baking updates into the security fabric.

The growth of Internet of Things (IoT) devices, estimated to be in the billions by 2025, poses a unique update dilemma. These commonly neglected endpoints make themselves attractive targets for attackers, thus requiring special measures to secure them. Spanning smart thermostats to industrial sensors, IoT devices tend to run outdated firmware due to their long lifecycles and minimal user interaction, making them vulnerable to exploits that can spread through networks [8]. For instance, a 2020 botnet attack through IoT leveraged unpatched devices to launch massive DDoS campaigns, a threat that persists through 2025 as connected infrastructure grows. IoT updates in the future will require lightweight, over-the-air patches with minimal bandwidth and processing demands, a trend being driven by firms like Amazon and Tesla with their connected devices. But manufacturers need to put security ahead of cost-saving, a change that regulators might mandate as IoT breaches escalate. This pattern is essential as IoT becomes more embedded in life, requiring updates that are commensurate with their size and complexity so that they do not become the weak point in cyber defenses.

Quantum-resistant cryptography is emerging as a long-term necessity, fueled by the threat of quantum computing, which has the potential to crack existing encryption methods and necessitate a massive software patching effort to secure data in the post-quantum era. By 2030, quantum computers can potentially break RSA and ECC encryption, exposing unpatched systems to retroactive decryption of stolen data, a threat that The Hacker News reported as early as 2022 [14]. NIST is finalizing quantum-resistant algorithms like CRYSTALS-Kyber in 2025, which software will need to adopt through updates to remain secure [14]. This transition will be gradual, starting in high-security sectors like finance and government, but it must be prepared for now—unpatched systems will especially be vulnerable as quantum power becomes accessible. The challenge is formidable: it will take decades to move billions of devices to new crypto standards, and older systems can never be updated, so they remain in a constant risk pool. Quantum-resistant updates represent a future necessity, driving the industry to a paradigm-changing security revolution that starts with current update methods.

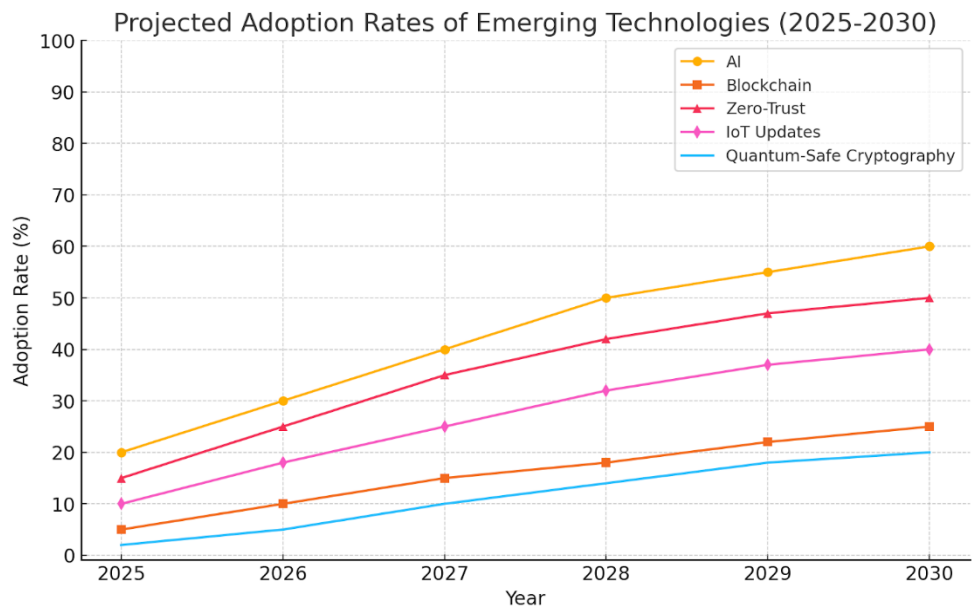

*Figure 1: Projected Adoption of Emerging Update Technologies (2025-2030)*

*Hypothetical data based on trends [8], [9], [10].*

This graph visualizes the gradual uptake of these technologies, highlighting their potential to transform update practices over the next decade.

## VI.    Legal and Regulatory Implications

The lack of updating software carries significant legal and regulatory implications that go beyond mere technical consequences, impacting organizations in financial, legal, and reputational terms—a side too often overlooked in the debate over cybersecurity until a breach forces it into the limelight. This section discusses the way compliance mandates and legal liability bring into sharp focus the necessity of regular updates in 2025. Regulatory frameworks now demand software updates as a minimum requirement for the protection of sensitive data, thereby putting organizations under the law to act or suffer extreme consequences that paralyze their operations. Regulations such as the Health Insurance Portability and Accountability Act (HIPAA) of the healthcare sector, the Gramm-Leach-Bliley Act (GLBA) of finance, and the Federal Information Security Management Act (FISMA) applicable to federal agencies in particular mandates that systems shall be maintained in an up-to-date state for protecting patient information, financial information, and national security data, respectively [10]. Failing to patch known vulnerabilities—like CVE-2023-34362 in MOVEit—can lead to audits, fines, or certification loss, as healthcare providers discovered after the 2023 breach when regulators reviewed their update practices [6]. In 2025, the regulations are tightening, with agencies like the U.S. Federal Trade Commission (FTC) calling for greater standards after breaches like Equifax, where the use of outdated software led to wide-scale data exposure [4]. Compliance is no longer optional; it is a necessity that ties updates to organizational viability, forcing even cash-poor organizations to patch or risk crippling penalties that could shut them down.

Aside from compliance, the legal implication of update negligence is lawsuits, fines, and reputational loss that could haunt organizations for years, escalating the cost of doing nothing in a litigious society. The 2017 Equifax breach is a textbook example: neglecting to patch Apache Struts cost $1.4 billion, encompassing class-action lawsuits by victimized consumers, regulatory fines by several jurisdictions, and settlements with state attorneys general, not to mention a stock plunge and customer flight that depleted its market position [4]. In 2025, this trend continues, with MOVEit victims being sued by clients whose data was stolen, arguing that slowness to update equated to negligence [11]. Loss of reputation contributes to these costs—companies lose trust, contracts, and partnerships, as Boeing lost when airlines questioned its trustworthiness after the 2023 attack [3]. Legal liability

2934

is growing, with courts increasingly holding companies accountable for preventable breaches, which makes updates a defense against both attackers and the law. Companies need to incorporate updates into risk management, consistent with regulations to maintain their reputation and avoid liability and maintain their position in a harsh digital and legal environment.

## VII.    Conclusion

Out-of-date software is a quiet yet formidable adversary in 2025, with convincing evidence—32% of attacks taking advantage of unpatched systems as reported by TechTarget—demonstrating its role as a keystone in the cyber security epidemic [1]. This paper has strictly analyzed this danger by using five case studies—MOVEit (2023), Boeing (2023), Equifax (2017), Log4Shell (2021), and MOVEit (2024)—each demonstrating how negligence turns vulnerabilities into breaches, with disastrous financial, operational, and societal impacts that echo through the years and across multiple industries [6], [3], [4], [15], [18]. Problems such as resistance by users, resource constraints, blind spots, patch fatigue, and third-party vulnerabilities continue to persist, as brought out by Gallagher Security and CompTIA, thereby forming a network of involved obstacles needing multi-pronged solutions [5], [10]. Automation, training, risk-based prioritization, centralized management, and best practices provide a holistic toolkit for overcoming these obstacles, as brought out by NCSC and Bitsight, thereby making updates an asset [7], [2]. Future directions—AI, blockchain, zero-trust, IoT upgrades, and quantum-resistant cryptography—promise to transform this landscape, with Figures 1 and 2 indicating the rising threat and emerging solutions [8], [9]. Legal imperatives, spanning from HIPAA compliance to Equifax-scale lawsuits, add urgency, tying upgrades to viability in an ever-more-regulated environment [10], [4]. To counteract these threats, organizations have to move swiftly: audit inventories to find out-of-date systems, install automation to patch in a timely manner, train users to change their mindset, and remain aware of threats and tech, making updates a foregone necessity in 2025's high-stakes cyber world.